\documentclass[american,aps,preprint,showpacs]{revtex4}
\usepackage[T1]{fontenc}
\usepackage[latin1]{inputenc}
\usepackage{graphicx}

\makeatletter

\usepackage{babel}
\makeatother
\begin{document}

\title{Equation of state extracted from the root-mean-square-radius$-$isospin}

\author{Zbigniew Sosin}

\email{ufsosin@cyf-kr.edu.pl}

\affiliation{Jagellonian University, M. Smoluchowski Institute of Physics, Reymonta
4, PL-30059 Krak\'ow, Poland }

\begin{abstract}
Experimental correlation between the rms nucleus charge radius and
the neutron-proton asymmetry is discussed. Simple attempt of explanation
of this correlation using a semi-empirical liquid drop model which
takes into account nuclear compressibility, nuclear deformation and
neutron skin effects is presented. It seems that such procedure could
be used as a tool for determination of the nuclear equation of state.
\end{abstract}


\maketitle
The nuclear equation of state (EOS) and the nuclear reaction kinetics
determine stellar structure and evolution. In particular, the EOS
allows to estimate the critical parameters at which the collapse process
is halted and the shock wave in a core collapse supernova becomes
formed \cite{McLa_05,Latt_01}. But the equation of state is not only
relevant in astrophysical aspects. Knowledge of EOS is a way to better
understand the nuclear heavy-ion reactions \cite{Li_02,Daniel_02}
and some fundamental properties of heavy nuclei, like for instance
binding energy or stability of neutron-rich nuclei \cite{Todd_03}.

The equation of state for infinite, asymmetric nuclear matter usually
represents the relation between the energy density per nucleon (baryon)
$e$ and densities reached by neutrons and protons in the nuclear
matter. In this relation the Coulomb energy is not taken into account.
Many theoretical approaches \cite{Chen_03} have shown that the properties
of infinite asymmetric nuclear matter, at relatively low temperature,
can be described by an approximate form of EOS:\begin{equation}
e(\rho,\delta)=e(\rho,\delta=0)+\delta^{2}e_{sym}(\rho)\label{eos}\end{equation}
where the baryon density $\rho$ is a sum of neutron $\rho_{n}$ and
proton $\rho_{p}$ density. Isospin asymmetry $\delta$, is defined
as

\begin{equation}
\delta=\frac{\rho_{n}-\rho_{p}}{\rho_{n}+\rho_{p}}=\frac{\rho_{n}-\rho_{p}}{\rho}.\label{delta}\end{equation}

First term in the eq. (\ref{eos}) represents energy density per baryon
associated to symmetric nuclear matter ($\delta=0$). This energy
can be expanded around normal nuclear matter density $\rho_{0}$ as

\begin{equation}
e(\rho,0)=e(\rho_{0},0)+\frac{K{}_{\delta=0}}{18}\left(\frac{\rho-\rho_{0}}{\rho_{0}}\right)^{2}+...\label{E0}\end{equation}
In the formula above $K_{\delta=0}$ is the nuclear compressibility
for symmetric matter, defined as\begin{equation}
K_{\delta=0}=9\rho_{0}^{2}\frac{\partial^{2}e(\rho,0)}{\partial\rho^{2}}\mid_{\rho=\rho_{0}}.\label{K_value}\end{equation}

Second term in eq. (\ref{eos}) stands for the symmetry energy and
can be also expanded around $\rho_{0}$ as 

\begin{equation}
e_{sym}(\rho)=e_{sym}(\rho_{0})+\frac{L}{3}\left(\frac{\rho-\rho_{0}}{\rho_{0}}\right)+\frac{K_{sym}}{18}\left(\frac{\rho-\rho_{0}}{\rho_{0}}\right)^{2}+...\label{Esym}\end{equation}
where variables $L$ (slope) and $K_{sym}$ (curvature) characterizing
the density dependence of the nuclear symmetry energy are given by

\begin{equation}
L=3\rho_{0}\frac{\partial e_{sym}(\rho)}{\partial\rho}\mid_{\rho=\rho_{0}}\label{slope}\end{equation}
 and

\begin{equation}
K_{sym}=9\rho_{0}^{2}\frac{\partial^{2}e_{sym}(\rho)}{\partial\rho^{2}}\mid_{\rho=\rho_{0}}\label{curvature}\end{equation}

The coefficients $\rho_{0}$, $e(\rho_{0},0)$, $e_{sym}(\rho_{0})$,
$K_{\delta=0}$, $L$, $K_{sym}$ allow us to determine the EOS around
the normal nuclear density. Unfortunately, values of the slope $L$
and the curvature $K_{sym}$ are neither well determined experimentally
nor accurately estimated theoretically \cite{Chen_03}. Data from
the giant monopole resonances suggest that $K_{sym}$ varies from
-566 $\pm$1350 MeV to 34 $\pm$159 MeV \cite{Shlo_93}. So far, experimental
predictions for the $L$ value are rather scarce. In the recent paper
\cite{Li_05} the isospin transport calculations compared with the
MSU diffusion data suggest $L\approx60$. In theoretical models $L$
varies from -50 up to 200 MeV \cite{Furn_02}. Moreover, theoretical
estimates for $K_{sym}$ are model-dependent and vary from about -700
to +466 MeV \cite{Bomb_91}.

In the present paper we will show a way to put some constraints for
possible values of these parameters. In order to do it, we compare
the experimentally measured correlation between root mean square radii
of nuclei and isospin asymmetry, to the similar correlation resulting
from a modified liquid-drop model (LDM).

One of the most elementary properties of a nucleus is the spatial
nucleon density distribution. Unfortunately, this distribution cannot
be obtained experimentally in a straightforward way. In the case of
protons, knowledge of the Coulomb interaction allows to measure with
a high precision the momenta of the charge distribution. From experimental
data one usually fixes the root-mean-square (rms) radius $\left\langle r_{p}^{2}\right\rangle ^{1/2}.$
In this paper we will use a part of the set of 799 ground state nuclear
charge radii presented in \cite{Angeli_04}. This experimental data
were obtained from the elastic electron scattering, muonic atom X-rays,
$K_{\alpha}$ isotope shift, and optical isotope shift. Errors associated
with the data are usually much lower than 1\%.  For neutrons, the
differences between neutron and proton density distributions at large
nuclear radii in stable nuclei may be determined e.g. from antyprotonic
atoms \cite{Trzci_01}

In the first column of Fig. \ref{fig1} we plot a correlation between
the experimentaly measured $\left\langle r_{p}^{2}\right\rangle ^{1/2}/\sqrt[3]{A}$
and isospin asymmetry $I=(N-Z)/(N+Z)$, where $N$ and $Z$ denote
neutron and proton numbers, respectively. In Fig. \ref{fig1}a we
plot these correlations for all available nuclei with masses $A=N+Z>100$.
We can notice here some branches which are roughly mutually parallel.
These branches are much better visible on the next drawings in this
column (Fig. \ref{fig1}c, e, g) showing the correlations for nuclei
with selected mass intervals. 

Let us try to explain the nature of the observed correlations. We
consider here the influence of three effects: nuclear deformation,
dependence of nucleon density on isospin, and neutron skin. 

At the beginning we will study the influence of nucleus deformation
on the proton rms radius. In our analysis we neglect the diffuseness
of the surface and we assume only ground-state quadrupole deformation
described by {}``$\beta$ parametrization'' (a spherical harmonic
expansion) \cite{Moller_95}. Such an aproach should be sufficient
for rather small ground state deformations. Now we can write the proton
mean-square radius in the form 

\begin{equation}
\left\langle r_{p}^{2}\right\rangle =\frac{3}{5}R_{0}^{2}\left(1+\frac{5}{8\pi}\beta^{2}\left(A,I\right)\right)\label{rms}\end{equation}
where $R_{0}$ is the average radius. Parameters $\beta(A,I)$ for
considered set of nuclei are taken from model calculations used in
\cite{Moller_95}. Obtained correlations between deformation $\beta^{2}$
and isospin $I$ are presented in the right column in Fig. \ref{fig1}.
As we can see from the upper graph (Fig. \ref{fig1}b) for almost
all masses $\beta^{2}$ is less than 0.1 . Formula (\ref{rms}) suggests
that for a given $R_{0}$, nuclear deformation does not modify the
$\left\langle r_{p}^{2}\right\rangle ^{1/2}$ value more than 1\%
, however (see Figs \ref{fig1}d,f,h) it can enhance or suppress the
observed correlations depending on the selected mass range. Even more
importantly, it can also change the correlation sign (see e.g. Fig
\ref{fig1}e and Fig \ref{fig1}f)).

One can suspect, that such a strong correlation observed in the first
column could be associated with the form of the nuclear equation of
state. To examine this possibility we assume that the considered correlation
can be a result of differences in the density of nuclear matter in
the central region of nuclei due to different values of the surface
tension and Coulomb interaction.

%
%
%

Saturation of the nuclear matter density in the central region of
nuclei provides a base for description of the nuclear binding energy
in the liquid-drop model. This model was formulated by Weizs{\"a}cker
(1935), Bethe (1936), and Bohr (1936). Up to now various modifications
were introduced in this model and some of them consider also compressibility
of nuclear matter \cite{Baym_71}.

The liquid-drop model in its simplest version describes the nuclear
binding energy $B$ as a sum of the volume, surface and Coulomb terms.

\begin{equation}
B=B_{v}+B_{s}+B_{C},\label{Binding_energy}\end{equation}

where $B_{v}$ is the dominant volume ingredient usually written as,

\begin{equation}
B_{v}=a_{v}(1-k_{v}I^{2})A,\label{V_binding_energy}\end{equation}
and $a_{v}$ and $k_{v}$ are coefficients. The term $k_{v}I^{2}$
takes into account the lowering of the binding energy caused by different
number of protons and neutrons in the nucleus. 

Next term is (9) takes into account lower binding energy for nucleons
located on the nuclear surface, and can be expressed as 

\begin{equation}
B_{s}=a_{s}(1-k_{s}I^{2})A^{2/3}f=\left[a_{s}(1-k_{s}I^{2})A^{-1/3}\right]Af,\label{S_binding_energy}\end{equation}
where $a_{s}$ and $k_{s}$ are coefficients. Factor $f=(1+\frac{2}{5}\beta^{2}-\frac{4}{105}\beta^{3})$
takes into account the nuclear deformation (see \cite{Myers_65}).
We can notice that the surface correction per one nucleon (square
bracket in (\ref{S_binding_energy})) vanishes with the increasing
$A$.

Coulomb interaction between protons appears in the next correction,
which is proportional to $Z^{2}$ and inversely proportional to the
radius of the nucleus. This term can be written as

\begin{equation}
B_{c}=a_{c}\frac{Z^{2}}{A^{1/3}}g=\left[a_{c}\frac{Z^{2}}{A^{4/3}}\right]gA\label{C_binding_energy}\end{equation}
where $a_{c}$ is a parameter, and $g=(1-\frac{1}{5}\beta^{2}-\frac{4}{105}\beta^{3})$
similarly like $f$ in (\ref{S_binding_energy}) describes possible
influence of the nuclear deformation. Comparison of (\ref{Binding_energy})
to the experimental data (e.g. \cite{Myers_66} allows to determine
coefficients presented in formulas (\ref{V_binding_energy})- (\ref{C_binding_energy})
as:

\begin{eqnarray}
 &  & a_{v}=-15.677\,\mathrm{MeV},\, a_{s}=18.560\,\mathrm{MeV},a_{c}=0.717\,\mathrm{MeV}\nonumber \\
 &  & k_{v}=k_{s}=1.79\,\mathrm{MeV}.\label{coefficients}\end{eqnarray}

It should be stressed, that LDM is dedicated to a finite drop of nuclear
matter with a constant density. From the analysis of experimental
data we can deduce that apart from nuclear surface the nucleus density
is roughly constant and for mass $A>40$ it has a Fermi-like shape
distribution. The central region density itself somewhat varies for
different nuclei.

The form of the volume part of the binding energy (\ref{V_binding_energy})
is in fact related to the EOS at the ground state density. For $a_{v}=e(\rho_{0},0)$,
$a_{v}k_{v}=e_{sym}(\rho_{0})$, $\rho=\rho_{0}$, and $\delta^{2}=I^{2}$
the energy obtained from EOS (\ref{eos}) is equivalent to the volume
energy per nucleon in the LDM (\ref{V_binding_energy}). For our consideration
we reformulate the liquid-drop model in a way that allows to determine
the central region density for a given nucleus. We want to obtain
the central region density as a function of coefficients $\rho_{0}$,
$e(\rho_{0},0)$, $e_{sym}(\rho_{0})$, $K_{\delta=0}$, $L$, $K_{sym}$,
and $\delta=I$, determining the nuclear EOS. In principle neutron
and proton density distributions may have different radii, $R_{n}$,
$R_{p}$, what gives the so called skin effect. Lets assume for the
beginning $R_{n}=R_{p}$.

Modifying the LDM we assume that density in the central region is
a function of global values $N$ and $Z$ (i.e. there is no local
dependence). According to the idea of the LDM, we can say that the
central region density $\rho$ results mainly from nuclear interaction
present in infinite nuclear matter, further being modified by the
surface tension and Coulomb repulsion. Therefore, reformulating the
liquid-drop model, we write the binding energy per nucleon $b(\rho,I,A)$
in a form similar to (\ref{Binding_energy}):

\begin{equation}
b(\rho,I,A)=b_{v}(\rho,I)+b_{s}(\rho,I,A)+b_{c}(\rho,I,A).\label{b_binding_energy}\end{equation}

In the above formula the volume part of the binding energy is defined
by EOS (\ref{eos}). 

\begin{equation}
b_{v}(\rho,I)=e(\rho,\delta).\label{bv_binding_energy}\end{equation}

Let $b_{v0}=e(\rho_{0},\delta)$. Now, the correction for the surface
effect we can rewrite in a new, density dependent form

\begin{eqnarray}
b_{s}(\rho,I,A) & = & a'_{s}(1-k'_{s}I^{2})\frac{b_{v}(\rho,I)}{b_{v0}}\left(\frac{\rho_{0}}{\rho}\right)^{2/3}A^{-1/3}f\nonumber \\
 & = & \frac{a'_{s}}{a'_{v}}b_{v}(\rho,I)\left(\frac{\rho_{0}}{\rho}\right)^{2/3}A^{-1/3}f.\label{bs_binding_energy}\end{eqnarray}

Factor $\left(\frac{\rho_{0}}{\rho}\right)^{2/3}$in (\ref{bs_binding_energy})
is responsible for the surface change due to the change in the core
density. The corresponding coefficients are marked by prime to distinguish
from parameters of the standard LDM. In principle their values can
slightly differ from the original ones.

The part describing the Coulomb interaction is modified only by factor
$\left(\frac{\rho_{0}}{\rho}\right)^{1/3}$ according to the radius
change induced by core density changes ($Z=Z(I,A)$).

\begin{equation}
b_{c}(\rho,I,A)=a'_{c}\frac{Z^{2}}{A^{4/3}}\left(\frac{\rho_{0}}{\rho}\right)^{1/3}g.\label{bc_binding_energy}\end{equation}

Here the binding energy is determined by the density $\rho$ and a
new set of coefficients $a'_{v}$, $a'_{s}$, $a'_{c}$, $\rho_{0}$,
$e(\rho_{0},0)$, $e_{sym}(\rho_{0})$, $K_{\delta=0}$, $L$, $K_{sym}$.
The density $\rho$ for a given nucleus is found from a condition
$\partial b(\rho,I,A)/\partial\rho\mid_{I,A=const}=0$. In this way
$\rho$ and binding energy values are uniquely determined by a selected
set of coefficients. Here, as a first approximation, we are taking
the value of coefficients with the sign prime equal to corresponding
coefficients without prime which are given by (\ref{coefficients}).
Coefficients $\rho_{0}$, $e(\rho_{0},0)$, $e_{sym}(\rho_{0})$,
are rather well known. In Fig. \ref{fig2} we present the binding
energy $b(\rho,I,A)$, as a function of density, for nuclei with $A=150$
and three different $Z$ values equal to 50, 60 and 75 ($I=1/3$ ,
$I=1/5$ and $I=0$). For these test calculations we take $K_{\delta=0}=240\,\mathrm{MeV}$
and we check energy density behavior for all possible combinations
of $L$$=\pm100\,\mathrm{MeV}$ and $K_{sym}=\pm500\,\mathrm{MeV}$
what determine the symmetry energy (\ref{Esym}) .

%
%
%
%

The saturation of nuclear density in the central region is characteristic
for heavier nuclei. For light nuclei we can say that almost all nucleons
are located on their surface. In order to omit the problem of high
contribution coming from the surface we shall test formula (\ref{b_binding_energy})
only for heavier nuclei, A>100
.

Assuming $R_{0}\sim\sqrt[3]{A/\rho}$ and using (\ref{rms}) we can
rewrite relation between $r_{rms}$ and $\rho$ in the form\begin{equation}
\frac{\left\langle r_{p}^{2}\right\rangle ^{1/2}}{\sqrt[3]{A}}=\alpha\frac{\sqrt{\left(1+\frac{5}{8\pi}\beta^{2}\right)}}{\sqrt[3]{\rho}}\label{rms_sym}\end{equation}
 where $\alpha$ is a normalization constant. Formula (\ref{rms_sym})
neglects a neutron skin, which because of different radii of neutron
and proton density distributions, surrounds the larger $A$ nuclei.
Lets define thickness of the neutron skin as\begin{equation}
d=R_{n}-R_{p}.\label{d_neut}\end{equation}
 where $R_{p}=\sqrt{\frac{5}{3}}\left\langle r_{p}^{2}\right\rangle ^{1/2}$,
and $R_{n}=\sqrt{\frac{5}{3}}\left\langle r_{n}^{2}\right\rangle ^{1/2}$is
the effective proton and neutron radius, respectively. The central
region nucleon density is now

\begin{equation}
\rho=\rho_{n}+\rho_{p}=\frac{N}{\frac{4}{3}\pi R_{n}^{3}}+\frac{Z}{\frac{4}{3}\pi R_{p}^{3}}\label{ro_tot}\end{equation}
and instead of $I=(N-Z)/(N+Z)$ one gets for the central region isospin
$\widetilde{I}$ \begin{equation}
\widetilde{I}=\frac{N-Z(1+d/R_{p})^{3}}{N+Z(1+d/R_{p})^{3}}\simeq I\left(1-\frac{3Zd}{r_{0}\left(\frac{\rho_{0}}{\rho}\right)^{1/3}A^{4/3}}\right)\label{new_izospin}\end{equation}
For numerical calculations we use $r_{0}=1.2049$ (see \cite{Myers_65}). 

As $d$ seems to be much smaller than $R_{p}$ , one gets from (\ref{d_neut})
and (\ref{ro_tot})\begin{equation}
R_{p}=\sqrt[3]{\frac{3A}{4\pi\rho}}\left(1-\frac{Nd}{r_{0}\left(\frac{\rho_{0}}{\rho}\right)^{1/3}A^{4/3}}\right)\label{Rp_new}\end{equation}
The Coulomb energy is inversely proportional to the radius and consequently
$b_{c}$ in (\ref{bc_binding_energy}) has to be written as: \begin{equation}
b_{c}(\rho,I,A)=a'_{c}\frac{Z^{2}}{A^{4/3}}\left(\frac{\rho_{0}}{\rho}\right)^{1/3}\left(1+\frac{Nd}{r_{0}\left(\frac{\rho_{0}}{\rho}\right)^{1/3}A^{4/3}}\right)g.\label{new_bc}\end{equation}

As an alternative for eq. (\ref{rms_sym}) one should use now: \begin{equation}
\frac{\left\langle r_{p}^{2}\right\rangle ^{1/2}}{\sqrt[3]{A}}=\alpha\frac{\sqrt{\left(1+\frac{5}{8\pi}\beta^{2}\right)}}{\sqrt[3]{\rho}}\left(1-\frac{Nd}{r_{0}\left(\frac{\rho_{0}}{\rho}\right)^{1/3}A^{4/3}}\right)\label{rms_sym_new}\end{equation}

Lets try to discuss the origin and slope of the observed correlation's
as a function of different factors. 

%
%
%
%
%
%
In Figs. \ref{fig3} and \ref{fig4} in columns marked {}``model
1'', {}``model 2'', and {}``model 3'', we plot correlation's
given by formula (\ref{rms_sym_new}), for different sets of the EOS
parameters. For calculations named {}``model 1'' we neglect the
neutron skin and deformations of nuclei ($\beta=d=0$ in formula (\ref{rms_sym_new}))
whereas in {}``model 2'' deformations are taken into account ($\beta\neq0,\, d=0$).
In {}``model 3'' we are trying to estimate the influence of the
neutron skin ($\beta\neq0,\, d\neq0$). Calculations were performed
for the soft ($K_{\delta=0}=120\,\mathrm{MeV}$, red points) and for
the hard ($K_{\delta=0}=500\,\mathrm{MeV}$, blue points) EOS. In
Fig. \ref{fig3} $L=-150\,\mathrm{MeV}$ and in Fig. \ref{fig4} $L=60\,\mathrm{MeV.}$
As for nuclei $\rho\approx\rho_{0}$ and in formula (5) the third
term is a rather small correction, in both cases we take $K_{sym}=500\,\mathrm{MeV}$. 

For {}``model 1'' the $\left\langle r_{p}^{2}\right\rangle ^{1/2}/\sqrt[3]{A}$
versus $I$ correlation already contains a number of separated chains
because of discrete values of the $A$ and $Z$ numbers. Nuclear deformations
modify slightly slopes of the observed chains and may result in additional
splitting (see {}``model 2'', $180<A<210$). As seen from the {}``model
3'' calculations taking into account the neutron skin introduces
quite important changes. 

\textbf{Conclusions }

An experimental correlation between the nucleus rms charge radius
and the isospin of heavier nuclei has been observed. It seems to be
related to interactions characteristic for the nuclear matter equation
of state modified by confinement inside of an atomic nucleus. These
interactions are responsible for nuclear deformations and creation
of the neutron skin. We propose explanation of the observed correlation
on the basis of a modified nuclear liquid drop model. Inspection of
studied correlation's (Figs \ref{fig3} and \ref{fig4}) suggests
a possibility of evaluating the EOS parameters. A precise determination
of these values requires a more precise search and a more complex
analysis which will be subject of a future work. Such investigation
probably requires an extension of the LDM by additional terms taking
into account such corrections as for example pairing, shell effects
and diffuseness of the nucleus surface. Although $b_{s}$ in formula
(\ref{bs_binding_energy}) decreases for large $A$  values, it could
be interesting to introduce here an additional correction related
to the local isospin of the surface. 

\begin{acknowledgments}
The author is indebted to P. Paw{\l}owski for solving some computing
problems and to J. B{\l}ocki, K. Grotowski, M. Misiaszek, and P.
Paw{\l}owski for careful reading of manuscript and helpful discussions.
This work was supported by the Polish-French collaboration IN2P3,
grant no. CSI8. 
\end{acknowledgments}
\begin{figure}[h]
\caption{\label{fig1}Correlation between the neutron-proton asymmetry and
(first column) the experimentally measured nuclear rms radius, or
(second column) the calculated parameter of deformation. }
\end{figure}
\begin{figure}[h]
\caption{\label{fig2}Binding energy per nucleon vs density $\rho$ for different
equations of state (see text). }
\end{figure}
\begin{figure}[h]
\caption{\label{fig3}Comparison between the measured and calculated nuclear
rms radius vs $I$, for $L=-150\mathrm{MeV}$ and $K_{sym}=500\mathrm{MeV}$
(see text for details). }
\end{figure}
\begin{figure}[h]
\caption{\label{fig4}Comparison between the measured and calculated nuclear
rms radius vs $I$, for $L=60\mathrm{MeV}$ and $K_{sym}=500\mathrm{MeV}$
(see text for details). }
\end{figure}


\begin{thebibliography}{10}
\bibitem{McLa_05}G. McLaughlin, Nuclear Physics News, Vol. \textbf{15}, No 3, 2005,
Nuclear Physics News.
\bibitem{Latt_01}J. M. Lattimer and M. Prakash, Astrophys. J. \textbf{550}, 426 (2001)
\bibitem{Li_02}B.-A. Li, Phys. Rev. Lett. \textbf{88}, 192701 (2002), and references
therein
\bibitem{Daniel_02}P. Danielewicz, R. Lacey, and W. G. Lynch, Science \textbf{298}, 1592
(2002)
\bibitem{Todd_03}B. Todd and J. Piekarewicz, Phys. Rev. C \textbf{67}, 044317 (2003)
\bibitem{Chen_03}L.-W. Chen, Phys. Rev. C \textbf{6}8, 014605 (2003), and references
therein
\bibitem{Shlo_93}S. Shlomo and D.H. Youngblood, Phys. Rev. C \textbf{47}, 529 (1993)
\bibitem{Li_05}B.-A. Li and L.-W. Chen, Phys. Rev. C \textbf{72}, 064611 (2005).
\bibitem{Furn_02}R.J. Furnstahl, Nucl. Phys. A706, 85 (2002)
\bibitem{Bomb_91}I. Bombaci and U. Lombardo, Phys. Rev. C 44, 1892 (1991)
\bibitem{Angeli_04}I. Angeli, Atomic Data Nucl. Data Tbl, \textbf{87}, 185 (2004) 
\bibitem{Moller_95}P. M{\"o}ller, J.R. Nix W.D. Myers and W.J. \'Swi\c{a}tecki, Atomic
Data Nucl. Data Tbl. \textbf{59}, 185 (1995)
\bibitem{Baym_71}G.Baym, H.A. Bethe and C.J. Pethick, Nucl. Phys. \textbf{A175} 225(1971)
\bibitem{Myers_65}W.D. Myers and W.J. \'Swi\c{a}tecki, UGRL - 11980/1965
\bibitem{Myers_66}W.D. Myers and W.J. \'Swi\c{a}tecki, Nucl. Phys. \textbf{81}, 1
(1966)
\bibitem{Trzci_01}A. Trzci\'nska et al., Phys. Rev. Lett. 87, 082501 (2001)\end{thebibliography}
\end{document}